\documentclass[10pt]{article}
\usepackage{amsmath,amssymb,graphicx}
\usepackage{color}
\usepackage{graphicx}
\usepackage{algorithm}
\usepackage{algpseudocode}
\usepackage{authblk}
\advance\textheight by 4cm
\advance\textwidth by 4.6cm
\advance\topmargin by -2.5cm
\advance\oddsidemargin by -2.3cm
\advance\evensidemargin by -2.3cm
\newcommand{\be}{\begin{equation}}
\newcommand{\ee}{  \end{equation}}
\newcommand{\ba}{\begin{eqnarray}}
\newcommand{\ea}{  \end{eqnarray}}

\def\openone{{1\!\!1}}
\def\NN{{\cal N}}
\def\OO{{\cal O}}
\def\KK{{\cal K}}

\title{Circuit complexity and functionality: \\
a statistical thermodynamics perspective}

\date{}

\author[1]{Claudio Chamon}
\author[1]{Andrei E. Ruckenstein}
\author[2]{Eduardo R. Mucciolo}
\author[1]{Ran Canetti}

\affil[1]{Boston University}
\affil[2]{University of Central Florida}

\begin{document}

\maketitle

\begin{abstract}
  Circuit complexity, defined as the minimum circuit size required for
  implementing a particular Boolean computation, is a foundational
  concept in computer science. Determining circuit complexity is
  believed to be a hard computational
  problem~\cite{KC00,Ilango2023,Huang2023}. Recently, in the context
  of black holes, circuit complexity has been promoted to a physical
  property, wherein the growth of complexity is reflected in the time
  evolution of the Einstein-Rosen bridge (``wormhole'') connecting the
  two sides of an AdS ``eternal'' black
  hole~\cite{Susskind2018lectures}. Here we are motivated by an
  independent set of considerations and explore links between
  complexity and thermodynamics for {\it functionally-equivalent}
  circuits, making the physics-inspired approach relevant to real
  computational problems, for which functionality is the key element
  of interest. In particular, our thermodynamic framework provides a
  new perspective on the obfuscation of programs of arbitrary length
  -- an important problem in cryptography -- as thermalization through
  recursive mixing of neighboring sections of a circuit, which can be
  viewed as the mixing of two containers with ``gases of gates''. This
  recursive process equilibrates the average complexity and leads to
  the saturation of the circuit entropy, while preserving
  functionality of the overall circuit. The thermodynamic arguments
  hinge on ergodicity in the space of circuits which we conjecture is
  limited to disconnected ergodic sectors due to {\it
    fragmentation}. The notion of fragmentation has important
  implications for the problem of circuit obfuscation as it implies
  that there are circuits of same size and functionality that cannot
  be connected via a polynomial number of local moves. Furthermore, we
  argue that fragmentation is unavoidable unless the complexity
  classes NP and coNP coincide, a statement that implies the collapse
  of the polynomial hierarchy of computational complexity theory to
  its first level.
\end{abstract}

Here we propose a physics-inspired statistical mechanics
approach to the behavior of reversible computational circuits of a
given functionality, in which circuit complexity - the minimal number
of gates required to implement that functionality - is treated as a
thermodynamic variable. While over the past decade the connection
between computational complexity and thermodynamics has emerged in the
course of explorations of the quantum mechanics of
black-holes~\cite{Susskind2018lectures,Brown_Susskind_Zhao,Brown_Susskind},
our own interest in this connection was motivated by a completely
different and independent line of enquiry, namely one aimed at
employing statistical mechanics techniques used to diagnose
irreversibility and chaos in complex physical systems in the design of
cryptographic tools for protecting confidential ``data-in-use'' -- data
and any manipulations of this data -- from bounded
adversaries~\cite{cipher-paper,eoc}.

\parshape=0

By contrast to the discussion of black holes that focuses on the growth of complexity for very large
random circuits and generally ignores the contribution of distinct circuits implementing the same functionality, the statistical mechanics approach presented here treats both complexity {\it and}
functionality. Ultimately it is the circuit functionality - the
specific computation implemented by a given circuit - that is the
central object of interest in most computational problems. Our
framework is based on reversible computing, which can be implemented
either as permutations $P$ acting on the space of $2^n$ strings of $n$
bits, or as unitary transformations $U$ acting on the $d^n$
dimensional Hilbert space of $n$ qudits with local Hilbert space
dimension $d$. For concreteness, in the body of the paper we focus on
universal circuits implementing classical permutations for which the
counting is discrete. \footnote{We note that here we focus on even 
permutations in the alternating group $A_{2^n}$, 
which can be decomposed into products of universal 3-bit gates~\cite{Coppersmith-and-Grossman}.
The realization of odd permutations requires either one additional $n$-bit
gate or one ancilla bit. More generally, the group $S_{2^n}$ can be embedded in the alternating group
$A_{2^{n+1}}$, an embedding that formalizes the incorporation of ancillas, and which also suggests that, given a computational model, the thermodynamic discussion presented here for even permutation extends
to odd permutations without essential changes.}
Since practical implementations of quantum
computation also use discrete sets of universal quantum gates to
approximate continuum unitaries, the extension to quantum circuits is
natural and is presented in Section A of the Supplementary Information.

The first part of the paper focuses on thermodynamic arguments based
on a novel notion of ``circuit ergodicity,'' which implies that the
space of circuits of a given size and functionality is covered
uniformly and displays a ``thermodynamic equilibrium'' behavior
analogous to that derived from a microcanonical ensemble for physical
systems. Within the circuit-thermodynamics framework we establish that
there are exponentially many ways to express a given functionality --
a permutation $P$ -- in terms of reversible gates; and connect the
scaling behavior as a function of complexity of two seemingly
unrelated counting problems: (a) how many ${\cal N}$-gate circuits one
can write for a given functionality, and (b) how many distinct
functionalities there are for circuits with given complexity
${\cal K}$. The connection between these quantities is tied to the
finite compressibility of typical circuits, e.g., those with gates
drawn randomly from a given gate set. Finite circuit compressibility
corresponds to a linear growth of complexity with number of gates (up
to its maximum value exponential in $n$) with a slope less than unity,
behavior that emerges as a result of functionality degeneracies.

The circuit-thermodynamics framework has a natural application to the
cryptographic concept of {\it program obfuscation} - a central problem
in cryptography - which, in conjunction with other cryptographic
primitives, enables a tremendously powerful set of cryptographic
tools.  Program obfuscation is the process of ``randomizing'' a
program while preserving its functionality.  Specifically, given a
computer program $P$, the process of obfuscating $P$ generates another
program $P'$ that (a) has a comparable size and the same functionality
as $P$; and (b) $P'$ provides no information about $P$ to an adversary
with polynomial resources, except for its
functionality~\cite{eprint-crypto}.
Within circuit thermodynamics, the equilibrium state defined by a
microcanonical ensemble in which an exponentially large number of
$\NN$-gate circuits with the same functionality appear with equal
probability naturally realizes the obfuscation of every circuit in the
distribution.

On a coarse-grained scale consistent with a thermodynamic treatment,
the microcanonical equilibrium state is accomplished through a
functionality preserving thermodynamic mixing process connecting local
equilibrium states associated with ``mesoscopic" subcircuits of the
original circuit. This mixing process, implemented via the ``flow" of
gates and complexity across mesoscopic size subcircuits, is analogous
to the equilibration of a set of initially separated containers of gas
once the constriction between them is removed.  This novel pathway to
circuit obfuscation based on glueing together locally equilibrated
subcircuits, which is conceptually different from the algebraic
global-circuit approach of state of the art
schemes~\cite{GargGH0SW16,DBLP:journals/iacr/BitanskyV15,DBLP:journals/iacr/AnanthJS15,DBLP:conf/pkc/LinPST16},
has recently inspired a new paradigm for cryptographic investigations
of circuit obfuscation~\cite{eprint-crypto}.

The second part of the paper returns to the issue of ergodicity in the
space of circuits, a notion that we assumed in formulating our
thermodynamic approach. This requires defining a dynamics in the space
of circuits that enables transforming two circuits into one another
while preserving size and functionality. We define a set of dynamical
rules that we refer to as ``$k$-string" dynamics according to which
one replaces $k$-gate subcircuits by equivalent subcircuits of equal
size. The resulting dynamics conserves both the functionality and size
of the original circuit. We argue that, generically, such models lead
to fragmentation of the space of circuits into disconnected
sectors. Thus, ergodicity holds and the thermodynamic framework only
applies within each sector. This conclusion raises important questions
about circuit obfuscation, questions that are connected with
fundamental assumptions of computational complexity theory. Finally,
we propose that a natural mathematical framework for formalizing the
notions of circuit collisions, fragmentation of the space of circuits,
and circuit ergodicity is the word problem in geometric group
theory~\cite{word-problem,geometry}, a connection we plan to explore
in future work.

\vspace{.2cm}

\subsection*{Results}


\noindent{\bf Counting circuits, entropy inequalities, and the
  thermodynamics of circuit complexity:} As noted above, there are
multiple ways of writing the same permutation $P$ using reversible
gates; the number of ways depends on the gate set $G$ used. We define
the circuit entropy
\begin{align}
  {\cal S}(P,\NN) =  \log_2 \Omega(P,\NN)\;,
  \label{eq:def-entropy}
\end{align}
where $\Omega(P,\NN)$ is the number of circuits realizing a permutation
$P$ with exactly $\NN$ gates. The latter definition
  implies (a) the inequality $\NN \ge \KK (P)$, where $\KK (P)$ is the
  {\it circuit complexity}, i.e., the minimum number of gates required
  to implement the permutation $P$; and (b) the sum-rule
  $\sum_P\;\Omega(P,\NN) = |G|^\NN$, where $|G|$ denotes the
  cardinality of the gate set used in the implementation of $P$. \footnote{For 3-bit gates in $S_8$, 
  $|G|=8! \binom{n}{3}$ and for gates in the universal set (NOT, CNOT, Toffoli), 
  $|G|=n + n(n-1) + n(n-1)(n-2)/2$.} We expect that, as is the case in physical systems, our circuit thermodynamic description is universal, with details of the gate set only appearing in the values of (intensive) non-universal parameters. 

The above counting parallels that used in the formulation of the
microcanonical ensemble in statistical mechanics. In this setting,
both $\NN$ and the circuit functionality, i.e., the permutation $P$
implemented by the circuit, are ``conserved quantities''. Furthermore,
we assume that all circuits implementing $P$ with $\NN$ gates appear
with equal weight in the counting, a condition equivalent to the equal
probability of microstates in the microcanonical ensemble.

A number of inequalities follow from the definition of the circuit
entropy in ~\eqref{eq:def-entropy}. The simplest one,
\begin{align}
  {\cal S}(P_1,\NN_1)
  + {\cal S}(P_2,\NN_2)
  \le {\cal S}(P_1P_2,\NN_1+\NN_2)
  \;,
  \label{eq:inequality_S}
\end{align}
expresses the fact that there may be more ways of implementing the
product $P_1P_2$ than simply sequentially implementing $P_1$ and then
$P_2$. In parallel with the entropy inequality in ~\eqref{eq:inequality_S}, the circuit complexity
satisfies the opposite inequality,
\begin{align}
  \KK(P_1) + \KK(P_2) \ge \KK(P_1 P_2)
    \;,
  \label{eq:inequality_K}
\end{align}
which reflects the simple fact that there may be shorter circuits
implementing $P_1P_2$ than the concatenation of $P_1$ and $P_2$.

Using the inequality in ~\eqref{eq:inequality_S}, we can immediately derive a lower bound
on the entropy ${\cal S}(P,\NN)$ in terms of the circuit complexity
$\KK(P)$ of the permutation $P$:
\begin{align}
 {\cal S}(\openone,\NN-\KK(P)) + {\cal S}(P,\KK(P))
  \le {\cal S}(P,\NN)
  \;,
  \label{eq:S-kappa}
\end{align}
where $\openone$ denotes the identity permutation, which has zero
complexity, i.e., it can be expressed without using any gate. 
~\eqref{eq:S-kappa} can be replaced with a more useful set of bounds, namely:
\begin{align}
 \left[\NN-\KK(P)\right]\;\log_2|G|^{1/2} +  {\cal S}(P,\KK(P))
  &\le {\cal S}(P,\NN)\nonumber\\
  & < \NN\;\log_2|G|
  \;.
  \label{eq:S-kappa2}
\end{align}

The lower bound in ~\eqref{eq:S-kappa2} is comprised of two contributions: the first involves the ``free volume" $\NN-\KK(P)$, with the complexity $\KK(P)$ acting as
  an ``excluded volume'', and depends on $P$ only through the
  complexity $\KK(P)$. The explicit form of this term is derived from ${\cal S}(\openone,\NN)$ via two steps:
(a) expressing $\NN$ as
$\NN = \sum_\ell a_\ell \;2^\ell$, where $a_\ell=0,1$ are the binary
coefficients in the expansion of $\NN$ in base 2 (we shall assume that
$\NN$ is even); and (b) using ~(\ref{eq:inequality_S}) multiple
times. This leads to ${\cal S}(\openone,\NN)\ge \sum_\ell a_\ell \;{\cal
  S}(\openone,2^\ell) \ge \sum_\ell a_\ell \;2^{\ell-1}{\cal
  S}(\openone,2) = \frac{\NN}{2}{\cal S}(\openone,2) =
\frac{\NN}{2}\log_2|G|$, where we used that a two-gate identity can be
written as the product of any gate $g$ and its inverse $g^{-1}$. 
The second term on the left hand side of 
~\eqref{eq:S-kappa} is independent of $\NN$ and accounts
  for the number of different possible circuits implementing the
  permutation $P$ within the minimum possible size $\KK(P)$. A simple argument, detailed in Section B of the Supplementary Information for circuits comprising of $3$-bit gates in $S_8$, shows that this term satisfies a lower bound proportional to $\KK(P)$, which we posit to be a thermodynamic property, valid for any choice of gate set G. 

Finally, the last upper bound in ~\eqref{eq:S-kappa2} follows immediately from the sum rule $\sum_P\;\Omega(P,\NN) = |G|^\NN$, which implies
$\Omega(P,\NN) < |G|^\NN$ and ${\cal S}(P,\NN) < \NN\;\log_2|G|$. 

Equation (\ref{eq:S-kappa2}) and the discussion accompanying it
connect the microcanonical ensemble entropy to the circuit complexity
and provide the foundation to what we refer to as the {\it
  thermodynamics of circuit complexity}. \footnote{In analogy to
  statistical physics, by introducing weights for different gates one
  could have also constructed a canonical ensemble for circuits.} In
particular, the linear $\NN$ dependence of both lower and upper bounds
of ${\cal S}(P,\NN)$ in ~\eqref{eq:S-kappa2} implies that
${\cal S}(P,\NN)$ is extensive in $\NN$. Moreover, the assumed
extensivity in $\KK(P)$ of the lower bound on ${\cal S}(P,\KK(P))$
motivated in the Supplementary Information, together with the upper
bound ${\cal S}(P,\KK(P)) < (\log_2 |G|) \KK(P)$ obtained
from~\eqref{eq:S-kappa2} evaluated for $\NN=\KK(P)$ imply that
${\cal S}(P,\KK(P))$ is extensive in $\KK(P)$, a property we express
as
\begin{align}
  {\cal S}(P,\KK(P)\approx \gamma (P;G)\; \KK(P)\;,
  \label{extensivity_of_minimal_circuit_entropy}
\end{align}
where $\gamma (P;G)$ depends on the permutation $P$ and the gate set,
$G$, used in implementing the permutation $P$.
  
The main assumption of our approach, motivated by the above discussion, 
is that ${\cal S}(P,\NN)$ is an extensive function of both $\NN$ and $\KK(P)$ - an
assumption analogous to that used in the microcanonical ensemble derivation of thermodynamics
for physical systems in which the extensive thermodynamic quantities are the number of particles and the energy.  An important consequence of the extensivity of the entropy with the ``free volume" $\NN-\KK$ is that the thermodynamic ensemble of $\NN$-gate circuits contains a number of circuits $\Omega(P,\NN)$ implementing any permutation $P$ that scales exponentially with the
  $\NN-\KK(P)$, behavior we will use repeatedly in what follows.  
 
To finalize the formulation of circuit thermodynamics,
we introduce the function $\omega(\KK,\NN)$ which, by contrast to
  $\Omega(P,\NN)$, which counts the number of $\NN$-gate circuits
  of {\it fixed functionality},
  counts all $\NN$-gate circuits of {\it fixed
  complexity}, $\KK$:
\begin{align}
  \omega(\KK,\NN)
  &=
  \sum_P \delta_{\KK,\KK(P)}\;\Omega(P,\NN)
  \nonumber\\
  &=
  \sum_P \delta_{\KK,\KK(P)}\;2^{{\cal S}(P,\NN)}
  \nonumber\\
  &\equiv \nu(\KK)\;2^{\bar{\cal S}(\KK,\NN)}
  \;,
  \label{eq:relation-to-omega}
\end{align}
where $\nu(\KK)= \sum_P \delta_{\KK,\KK(P)}$ is a ``density of
states'' that counts the number of possible functionalities (i.e.,
permutations) implemented by circuits of a fixed complexity, $\KK$; and $\bar{\cal S}(\KK,\NN)$ defines
an ``annealed average" of ${\cal S}(\KK,\NN)$ over $P$.
In turn, this allows us to write:
\begin{align}
  \sigma(\KK,\NN) = \log_2 \omega(\KK, \NN)
  &\equiv
  \log_2 \nu(\KK) + \bar{\cal S}(\KK,\NN)\
  \label{eq:relation-to-entropy}
  \;.
\end{align}

The extensively of ${\cal S}(P,\NN)$ with $\NN$ and $\KK(P)$ implies that $\bar{\cal S}(\KK,\NN)$ is itself extensive in $\NN$ and $\KK$.  The function
$\omega(\KK,\NN) = 2^{\sigma (\KK,\NN)}$, which defines the circuit
complexity weight distribution for $\NN$-gate circuits, peaks at
the extremum of the function $\sigma (\KK,\NN)$:
\begin{align}
\label{eq:extremum}
\frac{\partial {\sigma (\KK,\NN)}}{\partial \KK} {\biggr\rvert}_{\NN} = 0 =\frac{\partial{\log_2 \nu(\KK)}}{\partial \KK} + \frac{\partial \bar{\cal S}(\KK,\NN)}{\partial \KK} {\biggr\rvert}_{\NN}
\;.
\end{align}
Since the entropy $\bar{\cal S}(\KK,\NN)$ is an extensive decreasing function of complexity, ${\partial \bar{\cal S}(\KK,\NN)}/{\partial \KK} |_{\NN}=-\beta (\kappa)$, where $\beta (\kappa)$ is a positive intensive function of $\kappa = \KK/\NN$. At the extremum, $\kappa =\kappa^*$, $\beta \equiv \beta (\kappa ^*)$, and thus ${\partial\;{\log_2 \nu(\KK)}}/{\partial \KK}\vert _{\KK ^*} =\beta$, implying that $\nu (\KK ^*) = 2^{\beta \KK ^*}$ increases exponentially with complexity, consistent with the counting estimates for the number of circuits with given complexity 
$\KK$~\cite{Susskind2018lectures}.  At the extremum, ~\eqref{eq:relation-to-entropy} assumes the form
\begin{align}
\sigma (\KK ^*,\NN) = \beta \; \KK^* + \bar{\cal S}(\KK ^*,\NN)\
\label{eq:sigma}
\;.
\end{align}
It is gratifying that ~\eqref{eq:sigma} can be recast in a form
familiar from the thermodynamics of physical systems: if we interpret
$\KK$ as the negative of the energy, $\mathcal{E}=-\KK$ (or
equivalently, $\epsilon =\mathcal{E}/\NN = -\kappa$) and
$\beta =1/{T}$ as the inverse of the temperature $T$,
~\eqref{eq:sigma} can be rewritten in terms of the equilibrium free
energy $F(T,\NN)$,
\begin{align}
- {T}\; {\sigma_e}(\mathcal{E},\NN) \equiv F({T},\NN) = \mathcal{E} - {T}\;{{\bar{\cal S}}_e}(\mathcal{E},\NN)\;,
\label{eq:free-energy2}
\end{align}
where subscripts $e$ represent the respective $\sigma$ and
$\bar{\cal S}$ functions evaluated at the corresponding negative
energies. Within this correspondence, the smallest (large negative)
energy state - corresponding to high complexity - would represent a
low entropy solid while the largest (zero) energy state -
corresponding to low complexity - would represent a high entropy gas.

The direct analogy with statistical mechanics can be exploited further
in the calculation of the probability distribution of complexities for
$\NN$-gate circuits:
\begin{align}
  P_{\NN}(\KK)
  =
  \frac{\omega(\KK,\NN)}{\sum_{\KK=0}^{\NN} \omega(\KK,\NN)}   \;,
  \label{eq:probability-complexity}
\end{align}
where the form of $\omega(\KK,\NN)$ can be obtained by expanding
$\sigma(\KK,\NN)= \log_2 \omega(\KK,\NN) = \beta \KK + \bar{\cal
  S}(\KK,\NN)$ to second order in
$\Delta\KK=(\KK-\KK^*)$, the departure from the solution of the
extremum condition:
\begin{align}
  \log_2 \omega(\KK,\NN)
  &=
  \bar{\cal S} (\KK^*,\NN)
  +\beta \KK^*
  \nonumber\\
  &\;\;+
  \frac{1}{2}\; {\Delta \KK}^2 \;
  \frac{\partial ^2 \bar{\cal S}(\KK,\NN)}{\partial \KK^2}{\biggr\rvert}_{\NN,\KK=\KK^*} +\cdots\
\label{eq:expansion}\;.
\end{align}
Using the extensivity of the entropy in $\KK$ and $\NN$ we can write
$\bar{\cal S}(\KK^*,\NN) = -\beta \KK^* - \beta \mu \NN +\cdots$,
where we have borrowed the statistical mechanics notation involving
the equilibrium ``chemical potential'' $\mu$. At the extremum, this
leads to the simplification of the first two terms in
~\eqref{eq:expansion} to
$\bar{\cal S}(\KK^*,\NN) +\beta \KK^* = - \beta \mu \NN$ from which
it then follows that:
\begin{align}
  \omega (\KK,\NN)
  &=2^{- \frac{1}{2 \NN T^2  c_{\NN}} {\Delta \KK}^2}\;
    2^{-\beta\mu \NN}\;.
\label{eq:omega}
\end{align}
Thus, the probability distribution in
~\eqref{eq:probability-complexity} is a Gaussian peaked at
$\KK^*=\kappa^*\NN$ {\it linear} in $\NN$ with a width
(root mean square deviation)
$\Delta_{\rm rms} \propto \sqrt{\NN T^2 c_{\NN}}$, where
$c_{\NN} = -{\partial \KK}/{\partial T}{|}_{\KK^*,\NN}$ is a positive
intensive quantity analogous to the specific heat in thermodynamics,
which, in physical systems measures the increase in energy induced by
an increase in temperature and controls energy fluctuations around the
thermal equilibrium state at temperature $T$.  
Finally, ~\eqref{eq:omega} and the sum
rule
$ \sum_P\;\Omega(P,\NN) = \sum_{\KK =0}^{\NN} \omega
(\KK,\NN)=|G|^\NN$ determines the leading behavior of the chemical
potential, $\mu = -T \log_2 |G|$.

It is important to stress that we expect that generic solutions of the
extremum condition for $\kappa ^*$, which
depend on the gate set through the value of $\beta$, are neither 0
nor 1 but lie in between, $0<\kappa^* < 1$. This expectation
underscore two conclusions that we reach by generic thermodynamic
arguments alone, namely that: (a) the average complexity grows
linearly with the depth of the circuit, recovering results obtained by
explicit calculation~\cite{Haferkamp2022,li2022short}; and more
specifically that (b) typical circuits display a finite circuit
compressibility with a compression factor $\eta = (1-\KK^*/\NN)$, with
$0 < \eta <1$. We stress that these results rely on an important and
non-trivial feature that emerges from the thermodynamic arguments at
the root of ~\eqref{eq:extremum}, namely the balance between two competing
effects: the exponential increase in the density of states
$\nu (\KK) \approx 2^{\beta \KK}$ and the decrease in the entropy
$\bar{\cal S}(\KK,\NN)$ with increasing $\KK$. We also note that,
while our extensivity assumption for $\log _2 \nu(\KK)$ and the linear
increase of the average complexity with circuit depth should hold up
to a maximum complexity exponential in $n$,
$\mathcal{\KK}_{\rm max}\sim n\; 2^n$, our focus is on polynomial (in
$n$) size circuits with $\NN >>n$~\footnote{The
    circuit entropy $\bar{\cal S}$ continues to grow linearly with $\NN$
    even when $\NN$ increases beyond the maximum circuit
    complexity. We note that, if growth of the Einstein-Rosen
    wormholes is unbounded, circuit entropy could describe that
    growth.}. In Section C of the Supplementary
  Information we present a model for the growth of complexity of a random circuit, which 
  accounts for functionality degeneracy, and recovers results obtained from the thermodynamic treatment that relies on the extensivity of $\bar{\cal S}(\KK,\NN)$ and 
  $\sigma(\KK,\NN)$ with $\NN$ and $\KK$

\begin{figure}[h]
  \centering
  \vspace{0.5cm}
  \includegraphics[angle=0,scale=0.35]{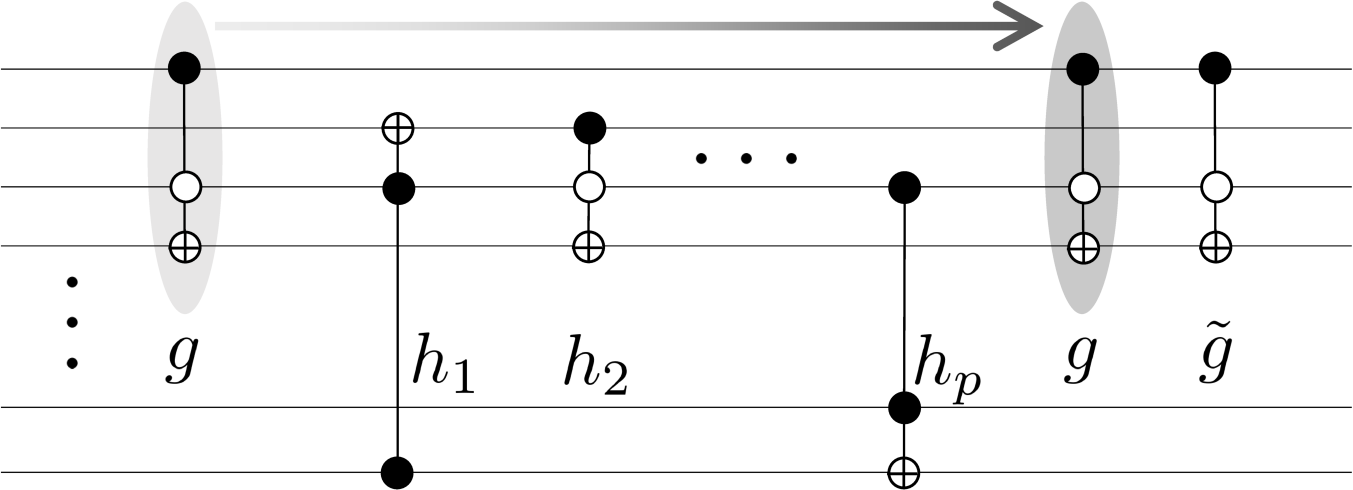}
  \vspace{0.5cm}
  \caption{An example of ``pushing'' a Toffoli gate $g$ past gates
    $h_1,h_2,\dots,h_p$ so as to annihilate $g$ with its inverse
    $\tilde g$ (identical to $g$ for a Toffoli gate), thereby reducing
    the circuit size by two gates. Typically, in a random circuit over
    $n$ bitlines, a Toffoli gate can travel past $\OO(n)$ gates with
    which it commutes before encountering either a gate with which it
    does not commute or, with probability $\OO(1/n^3)$, its
    inverse. The process depicted leads to a compression of the
    circuit size by a factor $(1-2\xi/n^2)$, where $\xi$ is a constant
    of $\OO(1)$.}
  \label{fig:collision}
\end{figure}

To further motivate the notion that generic circuits display a finite
compressibility with a compression factor $0 < \eta <1$, we consider
the following scenario leading to a lower bound on $\eta$. Consider a
random circuit of 3-bit Toffoli gates and imagine ``pushing'' a gate
through the circuit until the gate encounters either (i) a gate with
which it does not commute, in which case we stop; or (ii) its inverse,
in which case the pair (the gate and its inverse) annihilate,
decreasing the size of the circuit by two gates (see
Fig.~\ref{fig:collision}).\footnote{Note that here we are considering random circuits built gate-by-gate, as opposed to minimum circuits describing random permutations. The former are compressible while the latter are not.} For a Toffoli gate, the probability that it
does commute with a gate on its path is of the order $1-\OO(1/n)$, or
equivalently, a gate can be ``pushed'' through $\OO(n)$ gates before
either stopping as in case (i) or annihilating as in case (ii). The
overall probability of annihilation is $\OO(1/n^2)$, accounting for
the probability $\OO(1/n^3)$ that the inverse is met in any of the
$\OO(n)$ attempts before stopping. Hence, this process leads to a
compression of the circuit by a factor $(1-2\xi/n^2)$ of its original
size, where $\xi$ is a constant of $\OO(1)$. This implies that
circuits with greater than $\OO(n^2)$ universal (Toffoli) gates are
compressible with $\eta = 2\xi/n^2$, setting a lower bound for
compressibility of random circuits of universal gates. Indeed, since
the probability of annihilation of linear gates - NOTs and CNOTs -
scales more favorably as $1/n$ and $1/n^2$ respectively, circuits
comprised of gates from the universal set of Toffolis, CNOTs, and NOTs
are more compressible with a compression factor $\eta$ above the
Toffoli bound. [It is worth mentioning that since any linear circuit
can be implemented with $\OO(n^2)$ gates, purely linear circuits are
highly compressible, with $\eta \sim(1 - n^2/\NN)$.]

Finally, we note that, as already alluded to in the introduction, a
compression factor $0<\eta<1$ also implies that functionality
degeneracies are non-negligible, a conclusion which one could have
reached already from Eqs.~\eqref{eq:S-kappa} and ~\eqref{eq:S-kappa2},
which highlight the fact that there are exponentially many circuits
that realize any permutation $P$.


\vspace{.2 cm}
\noindent {\bf Thermodynamic mixing of circuits - an application to circuit obfuscation:}
As we have learned from physical systems, thermodynamics provides a platform for harnessing heat into useful work or 
for analyzing thermodynamic processes that transfer heat/entropy or molecular species. We find it reassuring that, in the context of circuits, this simplest of thermodynamic concepts - the entropy of mixing - provides a potentially important application to circuit obfuscation. As already alluded to in the introduction, obfuscation of circuits follows immediately from two tenets of circuit thermodynamics: (i) the fact that there are exponentially many comparable size circuits which implement the same functionality; and (ii) that in the fully mixed maximum entropy state for an ensemble of such circuits is described by a uniform (microcanonical) distribution, i.e., drawing a circuit from this distribution results in any one of the exponentially many $\Omega (P,\NN)$ realizations of the $\NN$-gate circuit with functionality $P$ with equal probability. Thus, any two different circuits of the same size and functionality drawn from the microcanonical ensemble of circuits cannot be distinguished from one another by an adversary with polynomial resources - the defining condition for the concept of circuit obfuscation.

Given an $\NN$-gate circuit of a given functionality, one can ask: how can one iteratively randomizes the gate makeup of a circuit while preserving its size and functionality in order to realize circuit obfuscation in the sense defined above? Strictly speaking, the microcanonical assumption, namely that all $\NN$-gate circuits of a given functionality $P$ appear with equal weight in the count $\Omega(P,\NN)$, hinges on ``ergodicity'' in the space
of circuits of a given functionality, a concept which implies a
``microscopic'' dynamical process and ``equilibration'' of collections
of gates in a circuit, analogous to the thermalization induced by
microscopic collisions of atoms or molecules in a gas. We shall return
to the complex and interesting question of microscopic dynamics
below. Here we concentrate on a coarse-grained model for thermodynamic
mixing on a ``macroscopic'' scale. We will imagine that by some process, which we will outline in the next section of the paper, we can divide the circuit into $M$ smaller ``mesoscopic" size subcircuits, and that we are able to fully equilibrate each of these subcircuits. Each of these ``mesoscopic'' subcircuits are assumed to
be large enough to obey the laws of circuit thermodynamics introduced
above but small enough so that an appropriate set of ``microscopic''
dynamical rules leads to equilibration in a time $\tau_{eq}$.

\begin{figure*}[h!]
  \centering
  \includegraphics[angle=0,scale=0.35]{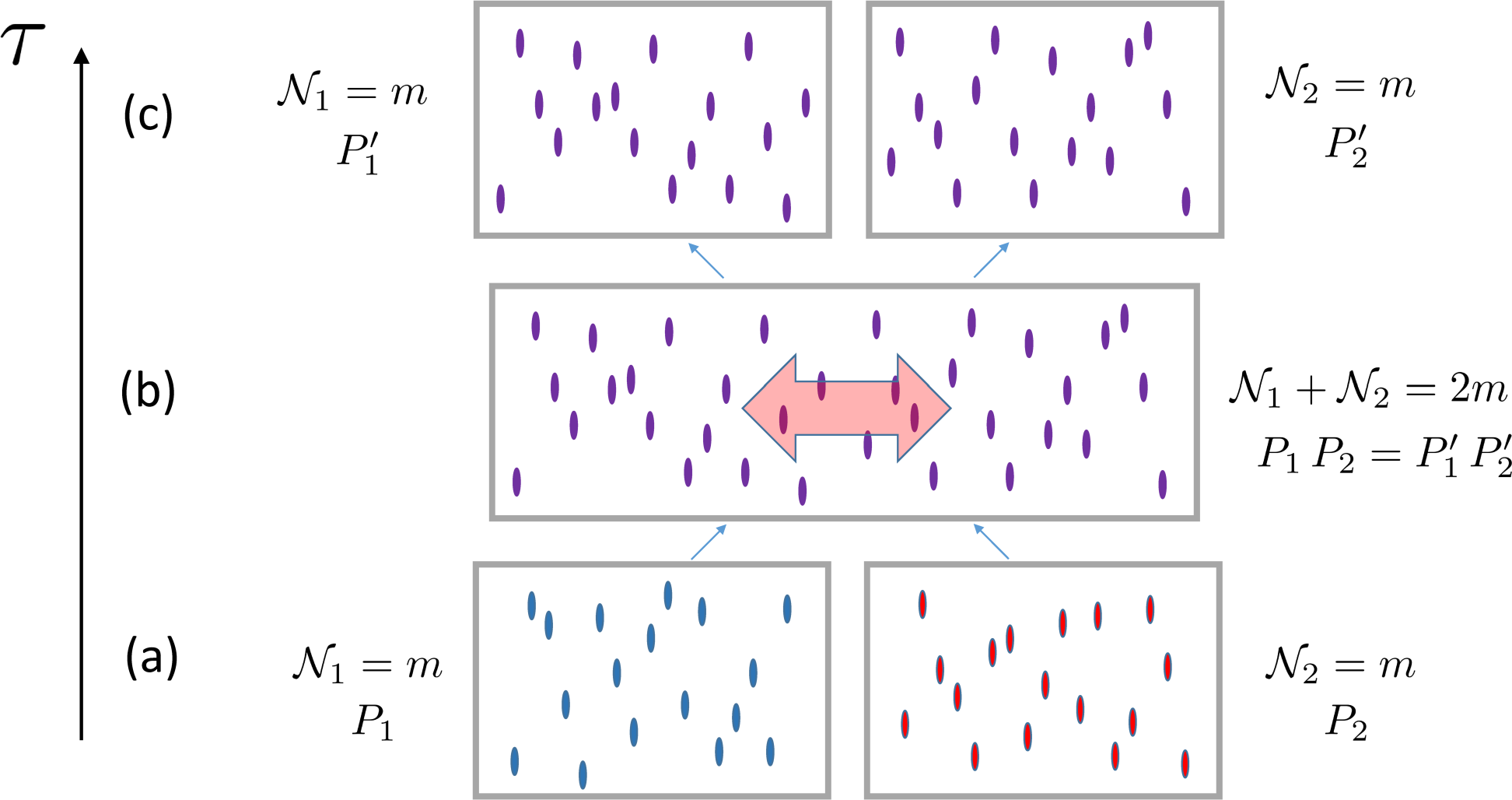}
  \caption{The elementary C-gate represents the equilibration between
    two circuits of equal sizes $\NN_1=\NN_2=m$. The two individual
    $m$-gate circuits, depicted in (a), have functionalities $P_1$ and
    $P_2$, respectively. In (b) they are brought into contact and
    exchange gates and functionality, realizing a combined
    functionality $P_1P_2$, while preserving the total number of gates
    $\NN_1+\NN_2=2m$. In (c), following equilibration (symbolized by
    the red double-arrow in (b)) the $2m$-gate circuit is split in the
    middle such that each of the partitions contains $m$ gates and
    represents, respectively, functionalities $P'_1$ and $P'_2$, with
    $P'_1\,P'_2=P_1\,P_2$. The thermodynamic process defining the
    C-gate increases the entropy.}
  \label{fig:C-gate}
\end{figure*}

The equilibration of an arbitrary (polynomial
size) circuit is then established by connecting a string of short $m$-gate
segments/subcircuits,  
More precisely, consider the situation depicted in
Fig.~\ref{fig:C-gate}, in which two $m$-gate subcircuits of
functionality $P_1$ and $P_2$, respectively, are allowed to exchange
gates and functionality via some dynamical rules. Thermalization at
the mesoscopic scale implies that, after a time scale $\tau_{eq}$, (a)
the counting of individual subcircuits (Fig.~\ref{fig:C-gate}a)
satisfy the microcanonical assumption, and (b) the concatenated
circuit with $2m$ gates and functionality $P_1 P_2$ satisfies the
thermodynamic inequalities
Eqs.~(\ref{eq:inequality_S})~and~(\ref{eq:inequality_K}).

\begin{figure*}[b!]
  \centering
  \includegraphics[angle=0,scale=0.35]{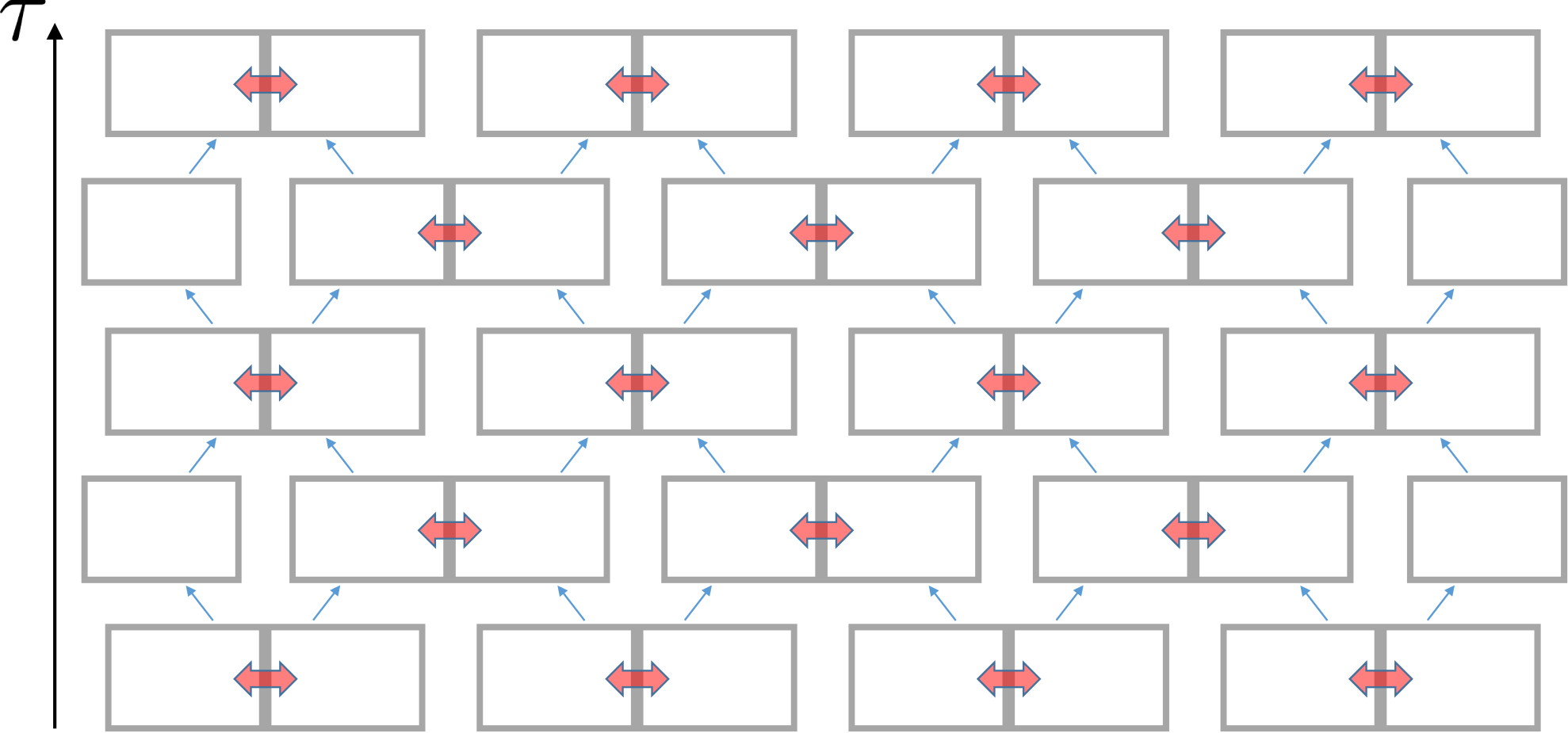}
  \caption{A brickwall C-circuit that progressively expands the
    ``mesoscopic" (local) equilibrium established between pairs of
    $m$-gate subcircuits (depicted by the gray boxes) into a
    thermodynamically mixed equilibrium state for the full circuit,
    while preserving the functionality of the original concatenation
    of gates. Neighboring $m$-gate subcircuits are brought into local
    thermal equilibrium via the exchange (depicted by the two-headed
    red arrows) of gates and functionality (while preserving both the
    number of gates and the combined functionality of the
    subcircuits). A subcircuit is paired with either its neighbor to
    the left or to the right, alternating in each time step (following
    the pattern of blue arrows).}
  \label{fig:brickwall}
\end{figure*}

Given the mesoscopic thermalization assumption, which we will revisit below, we are now in position
to define a coarse-grained model for thermodynamic mixing that takes
as input a circuit $C$ that is split into $M$ subcircuits, each
comprising of $m=\NN/M$ gates: $C = c_1 c_2 ...c_M$. A subcircuit
$c_i$ ($i=1,\dots,M$) can be thought of as a degree of freedom in a
$d$-dimensional space with $d=|G|^{m}$ states, i.e., a dit; and thus a
circuit can be viewed as a string of $M$ such dits. The coarse-grained
mixing of the full circuit $C$ is implemented as a circuit acting on
dit-strings, i.e., a ``circuit acting on circuits'' - hereafter
referred to as a C-circuit - built out of gates acting on dits, i.e.,
``gates acting on circuits'' - referred to as
C-gates. Fig.~\ref{fig:brickwall} depicts a brickwall C-circuit of
C-gates acting on a pair of neighboring dits $c_{i-1}(\tau)$ and
$c_{i}(\tau)$ in layer $\tau$ and evolving them into $c_{i-1}(\tau+1)$
and $c_{i}(\tau+1)$ after one ``time'' step, i.e., into layer
$\tau +1$ of the C-circuit. The brickwall C-circuit is a scrambler of
circuits, much as usual brickwall circuits of dit- or qudit-gates are
scramblers of states of dits or qudits~(see for example
Ref.~\cite{Harrow_Mehraban2023} and references therein).

The action of an individual C-gate, which takes place on a time scale
longer than $\tau_{eq}$, is based on the mesoscopic equilibration
assumption, and is implemented in three steps that parallel those
depicted in Fig.~\ref{fig:C-gate}, as follows: (a) take
$c_{i-1}(\tau)$ and $c_{i}(\tau)$, with functionalities
$P(c_{i-1}(\tau))$ and $P(c_{i}(\tau))$; (b) draw a circuit
$c_{\rm aux}$ uniformly out of the $\Omega(P, 2m)$ $2m$-gate circuits
with functionality $P=P(c_{i-1}(\tau))\,P(c_{i}(\tau))$; and (c)
split the $2m$-gate circuit $c_{\rm aux}$ into two $m$-gate circuits
$c_{i-1}(\tau+1)$ and $c_{i}(\tau+1)$. We note that the action of a
C-gate on the two dits $c_{i-1}$ and $c_{i}$ preserves functionality
of the product of the two associated sub-circuits, a ``conservation
law'' that maintains the functionality of the overall circuit. (We
also note that the stochastic process defined above could be replaced
by the action of a C-circuit built from deterministic C-gates with
given substitution truth tables chosen randomly.)

As alluded to above, the action of the brickwall C-circuit
progressively expands the ``local'' equilibrium within each of the
$m$-gate subcircuits into an thermodynamically mixed equilibrium state
for a full circuit $C$ of any size $\NN$. As already alluded to above, the equilibration process
induced via the C-circuit is analogous to the equilibration of
connected thermodynamic systems (e.g., containers of gas molecules)
that were initially isolated from one another. More specifically, this
thermodynamic mixing of circuits reflects three properties of the
explicit C-circuit implementation:
(i) the circuit entropy after the application of each C-gate never
decreases, but increases or remains the same;
(ii) through subsequent layers of the scrambling process,
functionalities of individual subcircuits change but the functionality
of the overall concatenated circuit is preserved; and, most
importantly,
(iii) the thermalization in the space of dits defining the action of a
stochastic C-gate and the layer-by-layer evolution of the circuit
leads to the branching into a multitude of paths, which implies that
memory of the initial circuit is lost, i.e., the scrambling
process is irreversible.

Given the one-dimensional brickwall arrangement of gates acting on $M$
dits, such as the C-circuit in Fig.~\ref{fig:brickwall}, the number of
layers required for scrambling the initial dit-string (in our case the
initial circuit $C$) should scale as $M^\gamma$. In random circuits
acting on dits without conservation laws
$\gamma=1$~\cite{Hayden_Preskill,Brown_Fawzi2013,Brown_Fawzi2015,Harrow_Mehraban2023}
while in the presence of locally conserved quantities we expect
$\gamma \ge
2$~\cite{Rakovszky_etal,Khemani_etal,Hearth2023unitary}. For the case
of scrambling by C-circuits, a C-gate acting on two subcircuits of
functionalities $P_1$ and $P_2$, respectively, may change $P_1$ and
$P_2$ but preserves $P_1 P_2$. This more complicated ``multiplicative" (rather than additive)
conservation law has not yet been analyzed in detail but we expect
that both the saturation of the entropy to its maximum attainable
value and the state of uniform average complexity,
$\bar{\KK_i} = \KK(P)/M$, for each of the $M$ subcircuits of a
C-circuit $C$ are reached within a time polynomial in $M$.
The above arguments imply that once any two initially distinguishable circuits of the same size and functionality are 
processed through a polynomial number of layers of a C-circuit, they become indistinguishable from one another by any adversary with polynomial resources.  

We note that the general line of reasoning presented so far makes
certain assumptions, some of which will be challenged below. In
particular, the notion of fragmentation of the space of circuits of a
given size and functionality into disconnected sectors, which we
introduce shortly, will restrict the thermodynamic framework to
individual sectors. Clearly, fragmentation of the space of circuits
will require sharpening of our definition of circuit obfuscation.

 \vspace{.2 cm}

\noindent {\bf Microscopic dynamics of circuits:}
All thermodynamics-based arguments presented thus far rely on the
assumption of ergodicity, namely that some microscopic dynamical rules
that connect circuits of same size and functionality lead to a uniform
covering of the space of all such circuits. Moreover, we assumed that
equilibration across the space of circuits is achieved in polynomial
time, i.e., that, given the dynamical rules, connecting any two
circuits can be achieved with a number of steps that scales
polynomially in the number of gates, $\NN$. This assumption raises a
number of interesting and up to now unexplored questions.

\begin{figure}[h]
  \centering
  \vspace{0.5cm}
  \includegraphics[angle=0,scale=0.28]{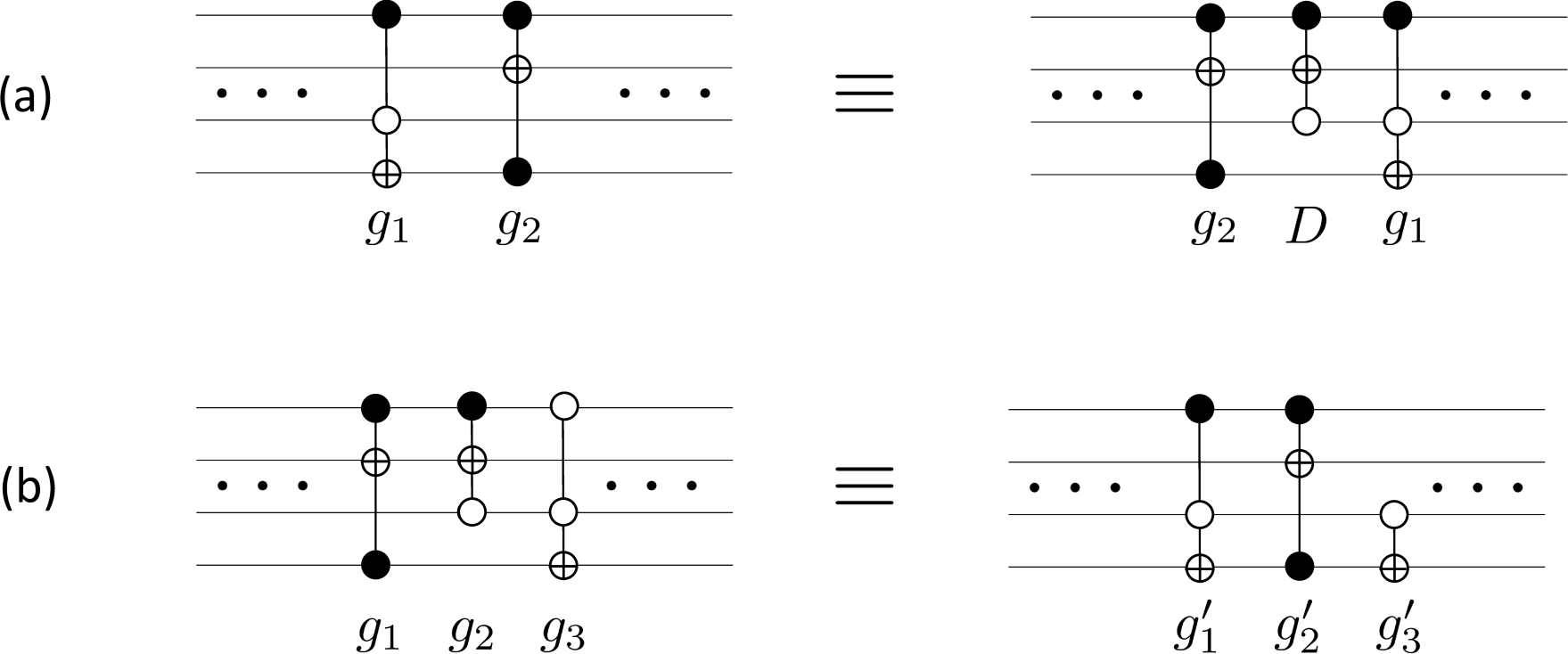}
  \vspace{0.5cm}
  \caption{(a) An example of a collision (the interchange)
    $g_1\,g_2\leftrightarrow g_2\, D\, g_1$, where the debris gate $D$
    is needed to preserve the functionality of the initial two-gate
    segment of the circuit. (b) An example of a substitution of a
    segment with $k=3$ gates,
    $g_1\,g_2\,g_3\leftrightarrow g'_1\,g'_2\,g'_3$. The same
    arrangements in (b) can be used as examples of two functionally
    equivalent circuits with $\NN=3$ that cannot be connected via a
    $k=2$ substitution rule.}
  \label{fig:subs}
\end{figure}

To begin with, by contrast to motion of molecules in a gas in the
course of collisions, which is governed by physical laws, there is no
unique or natural dynamics for moving/colliding gates in a circuit in
ways that preserve functionality. A naive notion of ``gate collisions",
analogous to collisions of gas particles, must take into account the
non-commutative algebra of gates in a universal set. If one defines a
gate collision as an interchange of gates $g_1$ and $g_2$ acting on shared
bitlines, then preserving functionality before and after the collision
implies, generically, a substitution
$g_1\,g_2\leftrightarrow g_2\, D\, g_1$, where $D$ is a ``debris''
gate needed so that algebraically $g_1\,g_2= g_2 \,D\, g_1$. An
example of such a collision is illustrated in Fig.~\ref{fig:subs}a for
two Toffoli gates. A macroscopic number of such debris-generating
collisions would inevitably lead to an irreversible increase in the
size of the circuit, violating the constraint of a fixed number of
gates.  

A more fruitful direction is to define a dynamics in the space of
circuits based on gate-substitution rules that exchange a string of
gates with an alternate string with same size and functionality. One
can view an $\NN$-gate circuit as a quasi-1D system, or a chain of
$\NN$ sites, in which a gate $g_i$ (a non-Abelian group element) is
placed at each site $i$. Global functionality is determined by
$P=g_1\;g_2\cdots g_\NN$, and a {\it local} microscopic dynamics must
preserve this overall functionality. The functionality-preserving
local dynamical model we have in mind involves the following
substitution of a string of $k$ consecutive gates:
\begin{subequations}
  \label{eq:sus-rules}
\begin{align}
\left(
g^{\,}_{i}, g^{\,}_{i+1},\dots,g^{\,}_{i+k-1}
\right)
&\longleftrightarrow
\left(
g'_{i}, g'_{i+1},\dots,g'_{i+k-1}
\right)
\\
g_{i}\,g_{i+1}\,\dots\,g_{i+k-1}
&\;\;=\;\;
g'_{i}\, g'_{i+1}\,\dots\,g'_{i+k-1}
\;.
\end{align}
\end{subequations}
An example of a $k=3$ circuit identity involving Toffoli and CNOT
gates is shown in Fig.~\ref{fig:subs}b. Substitution rules for fixed
(and small) $k$ can be built from a catalog of strings of $k$ gates
that multiply to the same permutation. Transition probabilities among
the $k$-length strings, in the case the catalog is exhaustive, can be
chosen to be
\begin{align}
&T_{
\left(
g^{\,}_{i},\dots,g^{\,}_{i+k-1}
\right)
,
\left(
g'_{i},\dots,g'_{i+k-1}
\right)
  }
\nonumber\\
&\hspace{1cm}
= \frac{1}{\Omega(g_{i}\,\dots\,g_{i+k-1}, k)}
\;
\delta_{g_{i}\,\dots\,g_{i+k-1}
,
g'_{i}\,\dots\,g'_{i+k-1}}
\;.
\end{align}
We note that the stochastic C-gate used above can be implemented via
such a transition matrix element with $k=2m$.  Alternatively, one can
dilute the connectivity associated with the $T$-matrix so that not all
pairs of ``$k$-strings" satisfying Eqs.~\eqref{eq:sus-rules} are
connected via a matrix element. We note that since the number of
circuits with $k$ gates is $|G|^k$, enumerating the equivalence rules
for large $k$ becomes prohibitive.

While to our knowledge this type of dynamical model has not been
discussed in the literature and a detailed study of the model is
outside the scope of this paper, we can already point to a set of
fundamental issues that have important implications for the discussion
of circuit thermodynamics.  In particular, our intuition suggests that
the space of circuits with functionality $P=g_1\;g_2\cdots g_\NN$
evolving via $k$-range rules will generically fragment into a number
of disconnected sectors. A simple example that supports the notion of
fragmentation is to consider a functionality-preserving dynamics that
only connects 2-strings if and only if two neighboring gates $g_{i}$
and $g_{i+1}$ commute, in which case we exchange
$(g_{i},g_{i+1})\leftrightarrow (g_{i+1},g_{i})$ with probability
1/2. This dynamics allows a gate to move left and right through the
list of gates in the circuit by passing other gates with which it
commutes, but not past those gates with which it does not
commute. This dynamics preserves the number of gates of each type in
the circuit and thus does not allow one to connect the two equivalent
circuits in Fig.~\ref{fig:subs}b. A less restricted dynamics with
$k=2$ in this same example would still not allow the two sequences of
three gates in Fig.~\ref{fig:subs}b to be connected.

Fragmentation implies that a particular dynamics is ergodic only
within individual sectors, and thus all the thermodynamic results
would apply, but only within each disconnected sector. Implicit in
this statement is that, within a given sector, ergodicity is reached
within a number of steps defining the particular dynamics that is
polynomial in $\NN$. In this case, the finite compressibility of
generic circuits is an example of a property that survives in the
fragmented system, where the compression factor should be determined
by some weighted average over fragments.

\vspace{.2 cm}
The intuition about fragmentation and polynomial thermalization is supported by computational complexity theory. Connecting two circuits of the same size and functionality by a polynomial number of local, functionality-preserving moves would imply NP = coNP, contradicting widely accepted beliefs of computational complexity theory. To understand this statement, recall that the complexity class NP contains YES decision problems that can be verified in polynomial time, while the class coNP contains problems for which NO solutions can be verified in polynomial time. If two circuits can be connected via a polynomial sequence of moves, Circuit Equivalence, a problem in coNP, is then also in NP because the sequence is itself a witness for the YES decision. Similarly, connecting two circuits through a polynomial sequence of moves implies that Circuit Inequivalence, a problem in NP, is also in coNP. Moreover, it is also well known that Circuit Equivalence and Circuit Inequivalence are among the hardest problems in their
respective coNP and NP classes, i.e., they are in classes
coNP-complete and NP-complete, respectively.  ``Completeness''
indicates that any problem in coNP or NP can be reduced, respectively,
to Circuit Equivalence or Circuit Inequivalence in polynomial
time. Given that a polynomial number of moves placed Circuit
Equivalence in NP and Circuit Inequivalence in coNP the conclusions of
the above line of argumentation is that NP = coNP, implying that fragmentation of the space of $\NN$ gate circuits implementing a permutation $P$ is inevitable, regardless of whether the sequence of moves is easy or hard to find.

Clearly, fragmentation and the accompanying broken ergodicity
significantly alters the IO construction presented above.  For a
system with multiple sectors, the relevant question becomes: given two
circuits $C_1$ and $C_2$, can one decide in polynomial time whether
they belong to the same ergodic (thermalized) sector or not?  Physical
intuition based on the scrambling of information, irreversibility, and
chaos in closed systems with large number of degrees of freedom leads
to a natural conjecture that, for non-trivial dynamical rules, this is
a hard (NP) decision problem.  If this is the case, then the
thermodynamic framework does in fact provide a path to
Indistinguishability Obfuscation of {\it any} two circuits, $C_1$ and
$C_2$. Otherwise the thermodynamic framework could only establish IO
for circuits in the same sector.



\vspace{.2 cm}

\subsection*{Discussion and future directions}

This paper presents a thermodynamic framework for describing
course-grained properties of large $\NN$-gate reversible classical (and quantum)
circuits with $\NN \gg n$ (with $\NN$ polynomial in $n$, the number of
bitlines of the circuit) and a given functionality, defined by the
permutation $P$ implemented by the circuit.  Our construction of
circuit thermodynamics is based on three assumptions that underpin the
logical consistency of the approach: (i) the functionality $P$ only
appears through the circuit complexity $\KK(P)$, i.e., the minimum
number of gates required for the implementation of the permutation
$P$; (ii) the entropy defined by counting of the number of possible
$\NN$-gate circuits implementing $P$ is extensive in $\NN$ and
$\KK(P)$; and (iii) ergodicity in the space of circuits, which as a
result of fragmentation can only occur in disconnected sectors,
requires a ``time" (i.e., number of dynamical moves) that is
polynomial in $\NN$, the size of the circuit.

The fragmentation of the space of circuits suggests a number of
questions we expect to address through more detailed analytical and
computational studies of the ``$k$-string" dynamics of reversible circuits: (i) is there is a critical value $k_c$ such
that if $k_c\le k \le \NN$ the space of circuits of size $\NN$ and
functionality $P$ becomes fully connected, and how does this value
scales with the number of bitlines $n$? (ii) If the space is
fragmented, how does the number of fragments scale with $k$ and $\NN$
and (possibly) the complexity of $P$? (iii) can one make more precise
statements about the hardness of deciding whether any two circuits
belong to the same or different sectors? We note that even though
we raise these questions in the context of the permutation group
$S_{2^n}$, the thermodynamic framework and the issues it raises can be
generalized to other groups.

In summary, the thermodynamic perspective to
  complexity and functionality of circuits proposed here charts a
  new line of inquiry. Developing this direction further and
  addressing quantitatively the questions it raises will require
  combining the intuition and tools of theoretical physics with
  mathematical ideas and techniques from the theory of classical and
  quantum computation. We believe that this work provides a framework
  that should stimulate new ways of thinking and fruitful interactions
  at the interface between physics and computer science. In
  particular, the issue of fragmentation which, to our knowledge, has
  not been explored by computer scientists, is a topic of much current
  interest to the physics communities working on classical and quantum
  dynamics~\cite{Ritort_Sollich2003,Moudgalya_etal2022,Khemani_etal2020,Moudgalya_Motrunich2022,Yang_etal2020}. Conversely,
  the problem of dynamics of systems with ``multiplicative'' rather
  than additive conservation laws (as is the case with the
  functionality of circuits) defines a new intriguing problem for
  theoretical physicists interested in classical and quantum dynamics.

We close by suggesting a direction that formalizes
  many of the ideas presented above, especially those related to the
  dynamics and ergodicity of systems of reversible circuits. The
  proposed direction starts from the observation that gate-based
  reversible classical and quantum computations are naturally framed
  in the language of group theory: gates represent group generators
  and the ``$k$-string'' substitution rules can be regarded as
  relations that define a {\it group presentation}. Thus, the
  ``$k$-string" dynamics defined in this work is intimately connected
  to the {\it word problem}~\cite{word-problem,geometry} of
  combinatorial and geometric group theory. This connection points to
  a formal mathematical framework that we expect will provide tools to
  further develop the circuit thermodynamics approach proposed here.

\vspace{.2 cm}

\subsection*{Acknowledgments}
We are grateful to Alexsey Khudorozhkov and Guilherme Delfino for
insightful discussions. This work was supported in part by NSF Grants
GCR-2428487 (A.E.R, C.C., R.C.) and GCR-2428488 (E.R.M), DOE Grant
DE-FG02-06ER46316 (C.C.) and a Grant from the Mass Tech Collaborative
Innovation Institute (A.E.R.). R.C., C.C., and A.E.R. also acknowledge
the Quantum Convergence Focused Research Program, funded by the Rafik
B. Hariri Institute at Boston University.


\newpage
\appendix

\section*{Supplementary Information} 



\subsection*{A - Extension to quantum circuits}
\label{sec:appendix-quantum}

The extension of the statistical thermodynamics approach to quantum
circuits can proceed via two routes that employ: (1) gates defined
over a continuous unitary group such as, for example, 2-qubit gates
drawn from a Haar-uniform measure on $U(4)$; or (2) a discrete set of
universal gates\footnote{The discrete gate sets are chosen so that the
  inverses of all gates in the set are also included.} such as CNOTs,
Hadamard gates H, and T gates ($\pi/8$ phase gates), or some other
discrete choice of an entangling 2-qubit gate and gates that
approximate arbitrary 1-qubit
rotations~\cite{brylinski2001universal}. For the purposes of this
discussion we choose option 2, which is both simpler and more relevant
to standard implementations of quantum computation.

We start by extending the expression for the number of circuits
realizing a permutation $P$ with $\NN$ gates drawn from a universal
set of classical reversible gates, $G$,
\begin{align}
  \Omega(P,\NN)
  =
  \sum_{\{g_1,g_2, \dots, g_\NN\}\in G^\NN}\;
  \delta_{P,g_1\,g_2\,\dots\,g_\NN}
  \;,
  \label{eq:def-Omega-P}
\end{align}
to the quantum case for which gates $u_1,u_2,\dots,u_\NN$ are drawn
from a discrete universal set of quantum gates, $G_Q$,
\begin{align}
  \Omega_{\epsilon}(U,\NN)
  =
  \sum_{\{u_1,u_2, \dots, u_\NN\}\in G_Q^\NN}\;
  \delta_\epsilon({U,u_1\,u_2\,\dots\,u_\NN})
  \;,
  \label{eq:def-Omega-U}
\end{align}
where
\begin{align}
  \delta_\epsilon(U,V)
  =
  \begin{cases}
    1, &\text{if } d(U,V)\le \epsilon\\
    0, &\text{if } d(U,V) > \epsilon
  \end{cases}
\;,
\end{align}
and $d(U,V)$ is a metric distance on the space of unitaries. The
distance between the two unitaries measures how well $V$ approximates
$U$, i.e., $d(U,V)$ quantifies the approximation error.  Approximating
a unitary as a product of gates in a discrete set is commonplace in
quantum computation~\cite{nielsen2002quantum}. Examples of metrics
quantifying the approximation error are
$d(U,V)=\underset{|\psi\rangle}{\max}||(U-V)\,|\psi\rangle||$ or
$d_{\rm tr}(U,V)= \text{arccos} \left(\frac{1}{D_H}
  |\text{tr}\,U^\dagger\,V|\right)$, where $D_H$ is the dimension of
the Hilbert space on which the unitaries act. (For example, the former
metric is used in Ref.~\cite{nielsen2002quantum} and the latter in
Ref.~\cite{Susskind2018lectures}.)  The metric satisfies the
inequality
\begin{align}
  d(U_1U_2,V_1V_2) \le d(U_1,V_1) + d(U_2,V_2)\;,
\end{align}
which leads us to infer that
$ \delta_{\epsilon_1}(U_1,V_1)\;\;\delta_{\epsilon_2}(U_2,V_2) = 1$
implies $\delta_{\epsilon_1+\epsilon_2}(U_1U_2,V_1V_2) = 1$ (the
reverse is not necessarily true). Using these relations we can write
\begin{align}
  \Omega_{\epsilon_1}(U_1,\NN_1)\;\Omega_{\epsilon_2}(U_2,\NN_2)
  \le
  \Omega_{\epsilon_1+\epsilon_2}(U_1U_2,\NN_1+\NN_2)
  \;.
  \label{eq:inequality-Omega-U}
\end{align}

We proceed by defining a quantity
\begin{align}
  \bar\Omega_{\delta}(U,\NN)\equiv \Omega_{\delta\NN}(U,\NN),
\end{align}
for which the choice of $\epsilon$ scales with $\NN$ via an
infinitesimal constant $\delta$, chosen so that the error
$\epsilon=\delta\, \NN$ is small for circuits of a bounded size (e.g.,
polynomial on the number $n$ of qubits).~\footnote{We note that the
  issues discussed above concerning the need of introducing the
  ``tolerance'', $\epsilon$, in the definition of $\Omega_\epsilon$ in
  ~\eqref{eq:def-Omega-U} has a direct analogue in physical
  systems.  Specifically, when defining a microcanonical potential for
  a physical system, $\Omega^{\rm ph}_\epsilon(E, N)$, which counts
  the number of microstates of given energy $E$ and particle number
  $N$, it is necessary to bracket states of energy between
  $E-\epsilon/2$ and $E+\epsilon/2$, defining an interval of size
  $\epsilon$ around the value of $E$. The inequality
  $\Omega^{\rm ph}_{\epsilon_1}(E_1, N_1)\;\Omega^{\rm
    ph}_{\epsilon_2}(E_2, N_2) \le \Omega^{\rm
    ph}_{\epsilon_1+\epsilon_2}(E_1+E_2, N_1+N_2)$ then follows.  Two
  comments are in order: first, notice that choosing
  $\epsilon=\delta\, N$ is natural as it keeps the intensive energy
  $E/N$ within a fixed interval of size $\delta$. Second, the actual
  choice of the $\epsilon$ is immaterial in statistical mechanics in
  the thermodynamic limit, because $\epsilon$ corresponds to the
  thickness of an energy shell in a $N$-dimensional space, and
  therefore the correction to the entropy from that shell is
  subextensive.  As a result, details on the specific choice of
  $\epsilon$ (beyond the definitional need to introduce it) are
  inconsequential.}

~\eqref{eq:inequality-Omega-U} then translates into an inequality
for $\bar\Omega_{\delta}$,
\begin{align}
  \bar\Omega_{\delta}(U_1,\NN_1)\;\bar\Omega_{\delta}(U_2,\NN_2)
  \le
  \bar\Omega_{\delta}(U_1U_2,\NN_1+\NN_2)
  \;,
  \label{eq:inequality-Omega-delta}
\end{align}
from which we derive the inequality for the entropy
${\cal S}_\delta(U,\NN) = \log_2 \bar\Omega_\delta(U,\NN)$, analogous
to that satisfied by the entropy permutation gates, namely,
\begin{align}
  {\cal S}_\delta(U_1,\NN_1)
  + {\cal S}_\delta(U_2,\NN_2)
  \le {\cal S}_\delta(U_1U_2,\NN_1+\NN_2)
  \;.
  \label{eq:inequality_S_delta}
\end{align}

We can similarly extend the defininion of complexity to account for
the approximation error:
\begin{align}
  \KK_\delta(U) \equiv
  \underset{\NN}{\min}
  \left\{\NN\;\; \Big|\;\; \bar\Omega_\delta(U,\NN) > 0 \right\}
  \;.
  \label{eq:K-delta-def}
\end{align}
It then follows from ~\eqref{eq:K-delta-def} and the inequality
~\eqref{eq:inequality-Omega-delta} that
\begin{align}
  \KK_{\delta}(U_1) + \KK_{\delta}(U_2)\ge \KK_{\delta}(U_1U_2)
  \;.
  \label{eq:K-inequality-K12}
\end{align}

With these definitions for $S_\delta(U,\NN)$ and $\KK_\delta(U)$,
which incorporate the approximation errors in realizing unitaries in
terms of a discrete set of universal quantum gates, the thermodynamic
approach for quantum circuits mirrors that presented in the body paper
for reversible classical circuits.

{\bf Microscopic dynamics of quantum circuits:} The local dynamical
rules introduced in the context of classical reversible circuits, can be easily extended to quantum circuits.
In the case of classical permutation gates, the functionality-preserving
local dynamical model involves substitution of strings of $k$
consecutive permutation gates (denoted by $g$),
\begin{subequations}
  \label{eq:sus-rules-sup}
\begin{align}
\left(
g^{\,}_{i}, g^{\,}_{i+1},\dots,g^{\,}_{i+k-1}
\right)
&\longleftrightarrow
\left(
g'_{i}, g'_{i+1},\dots,g'_{i+k-1}
\right)
\\
g_{i}\,g_{i+1}\,\dots\,g_{i+k-1}
&\;\;=\;\;
g'_{i}\, g'_{i+1}\,\dots\,g'_{i+k-1}
\;.
\end{align}
\end{subequations}
In the quantum case the rules can be generalized in a natural way to account for the approximation error in the substitution of strings of $k$ unitary gates (denoted
by $u$):
\begin{subequations}
  \label{eq:sus-rules-unitary}
\begin{align}
&\left(
u^{\,}_{i}, u^{\,}_{i+1},\dots,u^{\,}_{i+k-1}
\right)
\longleftrightarrow
\left(
u'_{i}, u'_{i+1},\dots,u'_{i+k-1}
\right)
  \\
  &\;\;\;\;\;\;
  d(u_{i}\,u_{i+1}\,\dots\,u_{i+k-1}
  \;,\;
u'_{i}\, u'_{i+1}\,\dots\,u'_{i+k-1}) \le \epsilon
\;.
\end{align}
\end{subequations}
The number of strings of $k$ gates satisfying the last condition is
given by $\Omega_\epsilon(\openone,2k)$.


\subsection*{B - Extensivity of circuit entropy at the complexity threshold}
\label{sec:appendix-extensitvity-with-Kappa}

Here we argue that the circuit entropy ${\cal S}(P,\KK(P))$ for
circuits of minimum length is extensive in the circuit complexity
$\KK(P)$. The upper bound ${\cal S}(P,\KK(P))\le \KK(P)\;\log_2 |G|$
derived from Eq. (5) of the main text establishes that the circuit entropy is at most extensive in
$\KK(P)$. Below we argue that ${\cal S}(P,\KK(P))$ is also bounded below by a term
proportional to $\KK(P)$. For concreteness, we use the gate set
$G$ comprised of 3-bit permutation gates in $S_8$, with $|G|=8! \binom{n}{3}$. We
consider circuits for which all inputs are affected by
at least one gate, i.e., none of the $n$ bitlines are untouched by the
circuit.

We proceed by considering the outputs of each of the $\KK(P)$ gates in the
circuit. Each of these outputs are either: (a) connected to the inputs of
another gate; or (b) exit unimpeded as output bitlines of the
circuit. In case (a) we have an internal link between the two gates. There
are $3\times\KK(P)$ outputs originating from the $\KK(P)$ 3-bit gates, which also include the exactly $n$ outputs of gates that connect all the way to the end of the circuit. As a result, the number of internal links 
within the circuit is given by $N_\ell = 3\times\KK(P)-n$. 
For each of these internal links one can negate the output of the gate
that acts before and the input of the gate that acts after, thus obtaining
different gates that are part of a circuit that yields the same
$P$. These negations can be implemented or not for each of the
internal links, which leads to a degeneracy of at least
$2^{N_\ell}$. For the polynomially-sized programs of interest in this paper, $\KK(P)\ge n$, and we thus obtain 
\begin{align}
S(P,\KK(P))\ge 3\KK(P)-n \ge 2\KK(P)
  \;.
\end{align}
Together with upper bound of Eq.~(5) in the main text, the above lower bound establishes the extensively of $S(P,\KK(P))$ with $\KK(P)$, which we express generically as,
\begin{align}
{\cal S}(P,\KK(P))\approx \gamma (P;G)\; \KK(P)
\;,
\end{align}
where the dependence on the specific permutation $P$ and the gate set
$G$ are made explicit in the coefficient $\gamma (P;G)$. In writing this expression we tacitly assume that
the argument given above will go through for any universal gate set G.

We note that the entropy computed above is associated with an internal redundancy that
arises through a ``gauge" transformation that is absorbed in
redefining two gates that share a link. 
Beyond this minimum set of redundancies, which occurs for any arrangement of gates, there are other ``structural" sources of entropy associated with different gate arrangements or circuit
architectures that yield functionally equivalent circuits.

\subsection*{C - Model for the dependence of the circuit entropy on the depth and complexity of computation}
\label{sec:walk}

In this section we present a derivation of the circuit entropy
$\sigma(\KK,\NN)$ by using a random walk model for the change in
complexity with the addition of gates, which should provide further
intuition about the discussion and results of the main text. This toy
model is based on the observation that, upon adding a single gate, the
circuit complexity increases or decreases by at most one. To see this,
consider the permutation obtained by adding a single gate $g$ to a
permutation $P$. Using the inequality for complexities, namely
$\KK(P_1P_2) \le \KK(P_1)+\KK(P_2)$, and the condition
$\KK(g)=\KK(g^{-1})=1$, we immediately obtain,
$\KK(Pg) \le \KK(P)+\KK(g) = \KK(P)+ 1$, and
$\KK(P) \le \KK(Pg)+\KK(g^{-1}) = \KK(Pg)+ 1$. Thus complexity can
only change by $\pm 1$ or $0$, as claimed, i.e.,
\begin{align}
  -1\le \KK(Pg)-\KK(P)\le +1
  \;.
\end{align}

Next consider $P=g_1\,g_2\,\dots\,g_{\NN}$, a circuit of $\NN$ gates,
and imagine the evolution of the partial complexity
$k = \KK(g_1\,g_2\,\dots\,g_n)$ after only the first $n\le \NN$ gates
are applied. The evolution of $k$ with $n$ is a walk that starts at
$(n,k)=(0,0)$ and ends at $(\NN,\KK)$, with the constraint that all
intermediate steps satisfy $k\ge 0$. 

For a random circuit, the evolution is a random walk that at each step
-- corresponding to the addition of one gate, or $\Delta n=1$ --
changes complexity by $\Delta k=\pm 1, 0$ with probabilities
$p_\pm,p_0$. During the initial growth of the circuit with $n$, i.e,
the early steps of the walk, we expect that these probabilities would
depend on history; however, we assume that $p_\pm,p_0$ reach a steady
state as the circuit length grows, with values that only depend on the
gate set $G$. We further posit that the number of (random) circuits of
$\NN$ gates and complexity $\KK$ is proportional to the number of
walks that start at $(n, k) = (0, 0)$ and end at $(\NN , \KK)$,
properly weighted by the probabilities $p_\pm$ and $p_0$. We proceed
with the computation of the number of such walks. 

Let $n_\pm$, $n_-$, and $n_0$ denote the number of steps involving
complexity changes by $\Delta k=\pm 1$ and $0$, respectively, which,
in turn satisfy the constraints $\NN=n_++n_-+n_0$ and
$\KK=n_+-n_-$. The number of such walks is given by
$\binom{\NN}{n_0}\;C(\NN-n_0, \KK)$, where the first factor counts the
number of ways of placing steps with $\Delta k=0$, and the second
factor counts the number of walks of $n_+$ ``up"-steps and $n_-$
``down"-steps with $n_++n_-=\NN-n_0$ that end at $\KK$ while never
crossing the $k=0$ line (since complexity cannot be negative).
$C(n,k)$ can be determined via a variant of the problem of computing
Catalan numbers (because of the restriction that $k\ge 0$ for all
intermediate steps); it satisfies the recursion
\begin{align}
  C (n, k) = C (n - 1, k - 1) + C (n - 1, k + 1)
  \;,
  \label{eq:recursion}
\end{align}
where $C (n, 0) = \binom{n}{n / 2} - \binom{n}{n / 2 + 1}$ is a
Catalan number that counts the number of balanced non-negative walks, i.e., walks that start and
end at $k=0$ after $n$ (even) steps but that can only explore points
with $k\ge 0$. 
By using Pascal's triangle relations it is then easy to see that the solution to the recursion relation of ~\eqref{eq:recursion} is given by
\begin{align}
  C (n, k) = \binom{n}{\frac{1}{2} (n+k)} - \binom{n}{\frac{1}{2} (n+k + 2)}
  \;.
  \label{eq:ansatz}
\end{align}
[For example, one can easily check that $C(k,k)=1$ consistent with the fact that there is only one walk that can reach
complexity $k$ after $k$ steps, i.e., after a walk with $n_+ =k$ and $n_-=0$.]

We can thus express the number of valid walks as
\begin{align}
  \binom{\NN}{n_0}\;C(\NN-n_0, \KK)
  =
  \frac{\NN!}{n_+! \, n_-!\,  n_0!}\;\;
  \frac{n_+-n_-+1}{n_+ +1}
  \;. 
\end{align}

We are now in position to compute the number of (random) circuits of $\NN$ gates of complexity $\KK$ by summing over the appropriately weighted walks that start at $(n,k)=(0,0)$ and end
at $(\NN,\KK)$:  
\begin{align}
  \omega(\KK,\NN)
  &=
  \frac{1}{Z}
    \sum_{n_0} \frac{\NN!}{n_+! \, n_-!\,  n_0!}\;\;
    \frac{n_+-n_-+1}{n_+ +1}
    \;\;p_+^{n_+}\,  p_-^{n_-}\,  p_0^{n_0}
    \;\; \delta_{\NN,n_++n_-+n_0}\;\delta_{\KK, n_+-n_-}
    \;,
    \label{eq:omega-sup}
\end{align}
where $Z$ is a normalization determined from the condition $\sum_\KK
\omega(\KK, \NN)=|G|^\NN$.

Let us consider the expression~\eqref{eq:omega-sup} with
$p_+\gg p_-,p_0$, a limit which is justified because for a generic circuit one expects that, upon addition a new (random) gate 
the complexity is much more likely to increase than to decrease or even
remain unchanged. Also, in this limit $n_+\gg n_-,n_0$ and $\frac{n_+-n_-+1}{n_+ +1}\to 1$, leading to the following expression for the normalization,
\begin{align}
  Z & \approx
      \frac{1}{|G|^\NN}
      \;\sum_{n_+, n_-, n_0} \frac{\NN!}{n_+! \, n_-!\,  n_0!}\;\;
      \;\;p_+^{n_+}\,  p_-^{n_-}\,  p_0^{n_0}
      \;\; \delta_{\NN,n_++n_-+n_0}
      \nonumber\\
    & \approx
      \frac{1}{|G|^\NN}\;(\;p_+ + p_- +  p_0)^\NN
  \;.
    \label{eq:Z}
\end{align}
We have not replaced $p_+ + p_- + p_0$ by unity in order to compute the expectation value as well as the fluctuations in the complexity $\KK$ from derivatives of $\ln Z$ with respect to the appropriate probabilities. More precisely,
\begin{align}
  \bar n_\pm = p_\pm\,\frac{\partial}{\partial p_\pm}\,\ln Z
  =
  \NN\,p_\pm
  \quad
  \text{and}
  \quad
  \bar n_0 = p_0\,\frac{\partial}{\partial p_0}\,\ln Z
  =
  \NN\,p_0
  \;,
\end{align}
from which it follows that the average complexity is given by:
\begin{align}
  \overline{\KK} = \bar n_+-\bar n_- = (p_+-p_-)\; \NN
  \;,
\end{align}
and the corresponding fluctuations in the complexity by:
\begin{align}
  \overline{\KK^2}-\overline{\KK}^2 = \left[p_+ (1-p_+) + p_-(1-p_-) + 2 p_+p_-\right]\; \NN
  \;.
\end{align}
These results recover the linear growth of the average complexity and the extensivity of the ``specific heat"  $ C_\NN = -({\partial \KK}/{\partial T}){|}_{\KK^*,\NN}$ controlling complexity fluctuations.

This toy model also allows us to recover the lower bound on the entropy of the identity permutation $\openone$, which corresponds to $\KK=0$ and for which
$\omega(0,\NN) = \Omega (P=\openone, \NN)$ and $\bar S(\KK(P)=0,\NN)= \sigma (\KK=0, \NN)$. The latter equality follows since there is only one functionality with $\KK=0$, namely the identity. Zero complexity corresponds to  $n_+=n_- = \frac{1}{2} (\NN-n_0)$,  in which case the expression for
$\omega(0,\NN)$ reduces to
\begin{align}
  \omega(0,\NN)
  &=
  \frac{1}{Z}
    \sum_{n_0} \binom{\NN}{n_0}\;C(\NN-n_0, 0)
    \;\;p_+^{(\NN-n_0)/2}\,  p_-^{(\NN-n_0)/2}\,  p_0^{n_0}
    \;.
    \label{eq:omega-zero}
\end{align}
Using the asymptotic limit of the Catalan number
$C(\NN-n_0, 0)\approx
\frac{2^{\NN-n_0}}{\sqrt{\pi/8}\;(\NN-n_0)^{3/2}}$ allows us to perform
the (binomial) sum over $n_0$, which yields
\begin{align}
  \omega(0,\NN)
  &=
    \frac{1}{Z}
    \;
    \eta (\NN)\; (2\,\sqrt{p_+\,p_-}+p_0)^\NN
    \nonumber\\
  &=
    \frac{1}{Z}
    \;
    \eta (\NN)\; [1-(\sqrt{p_+}-\sqrt{p_-})^2]^\NN
    \;,
    \label{eq:omega-zero-c}
\end{align}
where the factor $\eta(\NN)\approx 1/(\sqrt{\pi/8}\;\NN^{3/2})$ is only a polynomial correction.
Thus, up to a subextensive term in $\NN$,
\begin{align}
  \sigma(0,\NN) = \bar S(0,\NN)
  &=
    \NN\;\log |G| + \NN\;\log [1-(\sqrt{p_+}-\sqrt{p_-})^2]
    +
    \dots
    \;.
    \label{eq:entropies-zero}
\end{align}
The second term in~\eqref{eq:entropies-zero}, which depends on the probabilities, is negative
and thus one immediately obtains the upper bound
$\bar S(0,\NN)\le \NN\;\log |G|$. The lower bound for $\bar S(0,\NN)$ is derived by noticing that the probability that the
complexity decreases satisfies the bound $p_-\ge 1/|G|$, to account for
the fact that there is at least one gate in the set $G$ that, when
added, lowers the complexity, namely the inverse of the last gate
of the circuit (prior to the addition). That minimum value of
$p_-$ implies that $p_+\le 1-1/|G|$. We can thus bound the second
term in~\eqref{eq:entropies-zero}:
\begin{align}
  \NN\;\log [1-(\sqrt{p_+}-\sqrt{p_-})^2]
  &\ge
    \NN\;\log \left[1- \left(\sqrt{(1-1/|G|)}-\sqrt{1/|G|}\right)^2\right]
    \nonumber\\
  &\ge
    \NN\;\log [2\;\sqrt{(1-1/|G|)}\;\sqrt{1/|G|})]
    \nonumber\\
  &\ge
    -\frac{1}{2}\;\NN\;\log |G|
    \;.
\end{align}
\eqref{eq:entropies-zero} then reduces to:
$\bar S(0,\NN) \ge \NN\;\log |G| -\frac{1}{2}\;\NN\;\log |G|=\frac{1}{2} \;\NN\;\log |G|$, which recovers the lower bound for the entropy of circuits implementing the identity permutation
that we derived in the body of the paper by other means.


\newpage

\bibliographystyle{unsrt}
\bibliography{references}


\end{document}